\documentclass[sigconf]{acmart}
\usepackage{float}
\usepackage{titlesec}

\titlespacing*{\section}
{0pt}   % left indent
{0.5ex} % space before
{0.5ex} % space after

\titlespacing*{\subsection}
{0pt}
{0.5ex}
{0.5ex}

\titlespacing*{\subsubsection}
{0pt}
{0.5ex}
{0.5ex}

\usepackage{graphicx}  % for \rotatebox in Likert tables
\usepackage{array}     % for p{} columns
\usepackage{url}       % for \url{} in follow-up section
\newtoggle{comments} % this is like a toggle switch
\toggletrue{comments}
\usepackage{xcolor}
\usepackage[normalem]{ulem} % for \sout in \revdel (optional)

\newif\ifshowrevisions
\showrevisionsfalse % <- CLEAN VERSION

\ifshowrevisions
  \newcommand{\rev}[1]{\textcolor{blue}{#1}}
  \newcommand{\revdel}[1]{\textcolor{red}{\sout{#1}}}
\else
  \newcommand{\rev}[1]{#1}
  \newcommand{\revdel}[1]{}
\fi

\AtBeginDocument{%
  \providecommand\BibTeX{{%
    \normalfont B\kern-0.5em{\scshape i\kern-0.25em b}\kern-0.8em\TeX}}}

\copyrightyear{2026}
\acmYear{2026}
\setcopyright{cc}
\setcctype{by}
\acmConference[CSCW Companion '26]{Companion of the Computer-Supported Cooperative Work and Social Computing}{October 10--14, 2026}{Salt Lake City, UT, USA}
\acmBooktitle{Companion of the Computer-Supported Cooperative Work and Social Computing (CSCW Companion '26), October 10--14, 2026, Salt Lake City, UT, USA}
\acmDOI{10.1145/3785651.3831486}
\acmISBN{979-8-4007-2378-0/2026/10}

\usepackage{subcaption} % recommended (modern)

\usepackage{color}
\usepackage{enumitem}
\usepackage{ifthen}
\usepackage{ulem}
\usepackage{multirow}
\usepackage{siunitx}
\usepackage{breakcites} % added to prevent long citations
\usepackage{makecell}
\usepackage{multirow}

\definecolor{gray}{rgb}{0.8,0.8,0.8} 
\definecolor{green}{rgb}{0, 0.6, 0} 
\definecolor{orange}{rgb}{1, 0.5, 0} 	
\definecolor{mahogany}{rgb}{0.75, 0.25, 0.0}
\definecolor{purple}{rgb}{0.6, 0, 0.6}
\definecolor{darkgreen}{rgb}{0, 0.4, 0}
\definecolor{lightblue}{rgb}{0, 0.8, 1} 
\definecolor{lightgreen}{rgb}{0.9, 1, 0.9} 
\definecolor{lightred}{rgb}{1, 0.3, 0.3} 
\definecolor{navy}{rgb}{0, 0, 0.4}
\definecolor{highlight}{rgb}{1.0, 0.13, 0.32}
\definecolor{black}{rgb}{0.0, 0.0, 0.0}
\provideboolean{revising}
\setboolean{revising}{true}
\ifthenelse{\boolean{revising}}
{
	
    \newcommand{\delete}[1]{\sout{\textcolor{gray}{#1}}}

	\newcommand{\td}[1]{\textcolor{purple}{#1}} % Tawanna's comments
	\newcommand{\tv}[1]{\textcolor{orange}{#1}} % Tiffany's comments
	\newcommand{\ph}[1]{\textcolor{lightblue}{#1}} % Patrick's comments
	\newcommand{\vk}[1]{\textcolor{pink}{#1}} % Vaishnav's 
    \newcommand{\npar}[1]{\textcolor{mahogany}{<#1>}}
}{
	
	\newcommand{\delete}[1]{}

    \newcommand{\tc}[1]{}
        
	\newcommand{\td}[1]{}
	\newcommand{\tv}[1]{}  
	\newcommand{\jm}[1]{}
	\newcommand{\ph}[1]{}
	\newcommand{\vk}[1]{}
    \newcommand{\npar}[1]{} 
}

\usepackage{tokcycle}[2021-03-10]
\usepackage{xcolor}
\newcounter{wordcount}
\newcounter{lettercount}
\newcounter{wordlimit}
\newif\ifinword
\newif\ifrunningcount
\newif\ifsummarycount
\def\limitcolor{red}
\makeatletter
\newcommand\addtomacro[2]{\tc@defx#1{#1#2}}
\newcommand\changecolor[1]{\tctestifx{.#1}{}{\addcytoks{\color{#1}{}}%
  \tc@defx\currentcolor{#1}}}
\makeatother
\newcommand\dumpword{%
  \addcytoks[1]{\accumword}%
  \ifinword\stepcounter{wordcount}
    \ifrunningcount\addcytoks[x]{$^{\thewordcount,\thelettercount}$}\fi
    \ifnum\thewordcount=\value{wordlimit}\relax\changecolor{\limitcolor}\fi
  \fi%
  \inwordfalse
  \def\accumword{}}
\newcommand\addletter[1]{%
  \tctestifcatnx A#1{\stepcounter{lettercount}\inwordtrue}{\dumpword}%
  \addtomacro\accumword{#1}}
\xtokcycleenvironment\countem
  {\addletter{##1}}
  {\dumpword\groupedcytoks{\processtoks{##1}\dumpword\expandafter}\expandafter
    \changecolor\expandafter{\currentcolor}}
  {\dumpword\addcytoks{##1}}
  {\dumpword\addcytoks{##1}}
  {\stripgroupingtrue\def\accumword{}\def\currentcolor{.}
    \setcounter{wordcount}{0}\setcounter{lettercount}{0}}
  {\dumpword\ifsummarycount\tcafterenv{%
    \par(Wordcount=\thewordcount, Lettercount=\thelettercount)}\fi}

\begin{document}

\title{Co-Designing Community-Centered AI Education for Adults: A Midwestern Case Study}

\author{Yao Lyu}
\affiliation{%
  \institution{School of Information\\ University of Michigan}
  \city{Ann Arbor}
  \state{Michigan}
  \country{USA}}

\author{Leonymae Aumentado}
\affiliation{%
  \institution{University of Michigan}
  \city{Ann Arbor}
  \state{Michigan}
  \country{USA}}

\author{Holden Winton}
\affiliation{%
  \institution{Department of Sociology\\ University of Michigan}
  \city{Ann Arbor}
  \state{Michigan}
  \country{USA}}

\author{Jared Lee Katzman}
\affiliation{%
  \institution{School of Information\\ University of Michigan}
  \city{Ann Arbor}
  \state{Michigan}
  \country{USA}}

\author{Sparkle Berry}
\affiliation{%
  \institution{Eastside Community Network}
  \city{Detroit}
  \state{Michigan}
  \country{USA}}

\author{Zachary Rowe}
\affiliation{%
  \institution{Friends of Parkside}
  \city{Detroit}
  \state{Michigan}
  \country{USA}}

\author{Kimberly Sanders}
\affiliation{%
  \institution{University of Michigan}
  \city{Detroit}
  \state{Michigan}
  \country{USA}}

\author{Tawanna R. Dillahunt}
\affiliation{%
  \institution{School of Information\\ University of Michigan}
  \city{Ann Arbor}
  \state{Michigan}
  \country{USA}}

\renewcommand{\shortauthors}{Lyu, et al.}

\begin{abstract}
Artificial Intelligence (AI) education is increasingly important, yet adults outside higher education receive less attention. We report a case study of an AI education session with 54 adults (48 in-person and 6 virtual) in a predominantly African American community on the east side of a major Midwestern city. We ask: ``What does AI education for adults outside formal educational systems look like in practice?'' and ``What does this AI education session reveal about AI literacy at the community level?'' Through a co-designed session developed with community partners, we found that concerns about AI persisted but shifted to specific, locally grounded questions about AI design and deployment. We also discuss AI literacy from a community capacity perspective and argue for AI literacy frameworks grounded in local community contexts that strengthen community capacity.
\end{abstract}




\maketitle

\section{INTRODUCTION}

Artificial Intelligence (AI) promises to transform society by improving employment \cite{constantinides_ai_2025}, education \cite{roschelle_Benchmarks_2025}, and healthcare \cite{gupta_transforming_2024}. As AI becomes pervasive and fundamental in everyday life, AI education has become essential, not only for AI professionals but also for the public. Helping the broader public understand AI is becoming an increasingly significant topic. However, current AI education efforts cover an uneven population. According to a scoping literature review \cite{laupichler_artificial_2022}, most existing programs focus on students and professionals, paying less attention to adults outside educational systems. Moreover, most AI education frameworks tend to define AI literacy as technical competencies, productivity proficiencies, and professional skills \cite{ng_Conceptualizing_2021,ng_AI_2021,register_Learning_2020}, paying less attention to the situated and contextualized scenarios of AI education. Current CSCW studies have pointed out that, when understanding technology literacy and knowledge, educators and researchers need to take factors like social relationships and power dynamics into consideration \cite{ajmani_Whose_2024}, especially at the community level \cite{lee_Sociocultural_2025}. Therefore, there remain gaps in both the populations covered by AI education and the understanding of AI literacy at the community level. To build on this line of CSCW research and address these gaps, our work reports a case study of a co-designed AI education session for adults outside educational systems in a local community, guided by the following research questions:

RQ1: What does AI education for adults outside formal educational systems look like in practice?

RQ2: What does this AI education session reveal about AI literacy at the community level?

We explore these questions in a predominantly African American community on the east side of a major Midwestern city. We took a community-based approach and co-designed the AI education session with the community partners. We consider AI education not as a passive process of delivering knowledge, but as an interactive process of understanding one another \cite{dahl_facilitation_2025,hui_Was_2024,gautam_Dynamic_2024}. The education session emphasized accessibility, local culture, and conversational learning. \rev{A total of 54 adults attended the session, including 48 in person and 6 virtually. A total of 25 participants provided demographic information\footnote{Demographic questions were optional, as we wanted to make the session feel more like a community gathering than a formal training program}: 17 participants identified as female and 4 as male; 17 identified as African American or Black, with one participant additionally identifying as Black/Brown and another as African American/Irish; and attendees represented diverse educational backgrounds, including 10 with college-level education, 8 with some college experience, and 6 with a high school education or less. \rev{Additionally, 22 participants reported their ages, with 17 aged 55 or above.}}

By conducting pre- and post-surveys and interactive polling, we found that community members expressed concerns about AI, and \textbf{those concerns remained after the session but became more specific and actionable.} Inspired by prior work on community capacity \cite{chaskin_Building_2001,dillahunt_Development_2025}, we discuss AI literacy and argue for AI education approaches that are grounded in and designed for community capacity. \textbf{We emphasize that AI literacy should resist a generic, top-down framing and instead be grounded in community context, recognizing that community members themselves should hold the power to define what AI means, where it fits into their lives, and what it can do for them.} Our contributions include: (1) A community-informed approach for co-designing an AI education session, including key decisions made; (2) Empirical results outlining community attendees' initial feedback on the session, as well as their understanding, perceptions, concerns, and desires for AI technology; (3) Analytical discussion of understanding AI literacy from the community capacity perspective; and (4) Implications for future community-centered AI education efforts.
\section{RESEARCH CONTEXT AND COMMUNITY PARTNERSHIP}
The research site comprises two local community organizations on the east side of a major Midwestern city. The two organizations serve communities that have historically experienced economic and social instability, exclusion, and limited voice in technology deployment decision-making. The research team and the organizations have maintained a collaborative relationship with the communities for over a decade through community-based research projects on employment, entrepreneurship, and community digital capacity. As AI becomes increasingly prevalent, the research team and the community partners decided to work together to better understand AI's emerging role in society and within the community. Therefore, the research team and community partners envisioned an initial AI education session.

\section{CO-DESIGNING THE AI EDUCATION THROUGH A COMMUNITY-BASED APPROACH}

The research team and the community partners decided to co-design an AI education session. For the particular session, we set two primary goals: (1) To understand the baseline understanding of AI among community members, recognizing that community members more than likely had highly varied prior exposure to familiarity with AI. (2) To surface community members' interests, concerns, questions, and desires related to AI. \rev{The research team held three meetings with one or both community partner organizations between April and May 2025. All community organization partners were invited to each meeting. Community partners ensured that the session effectively served community members' interests (e.g., establishing a baseline understanding of AI) while also meeting the research purposes of the session (assessing community members’ current understanding of AI to inform planning for future community sessions). Community partners provided feedback on session creation and hosting, including making the session interactive, clarifying the definition of AI, providing examples of AI, hearing community members' voices on attitudes toward AI during the session, and ensuring accommodations. Community partners advertised the study via email and printed flyers. We obtained Institutional Review Board (IRB) approval from our university before reaching out to participants.}

\subsection{Co-creating the Education Session}

During this stage, the research team and the community partner organizations met regularly to discuss the content of the materials. The research team started by developing the slides and finding interactive activities based on feedback from community partners. The research team then presented the initial draft to the community partners for feedback. After several rounds of feedback and revision, the research team finalized the materials. The key decisions made (as well as the key values surfaced) during this stage include:

\textbf{Balanced Topics}. \rev{We selected a balanced set of topics for the session, informed by community partners' insights into what would be most relevant to community members.} Balanced topics equipped community members with a more comprehensive mental map of AI. To avoid misrepresentation of AI and its impact, our content included (1) a comprehensive set of AI topics covering both positive and negative sides of AI and (2) a series of prompt questions on attendees' various perceptions of AI, like desires, concerns, and interests. The balanced topics were driven by the understanding that community members could encounter AI in any scenario; therefore, the educational materials, like the topics, should reflect this variety. By presenting both the positive and negative sides of AI comprehensively, the session avoided influencing community members' own judgments of AI and coming across as overly negative or positive.

\begin{figure}[H]
    \centering
    \begin{subfigure}[t]{0.4\linewidth}
        \centering
        \includegraphics[width=\linewidth]{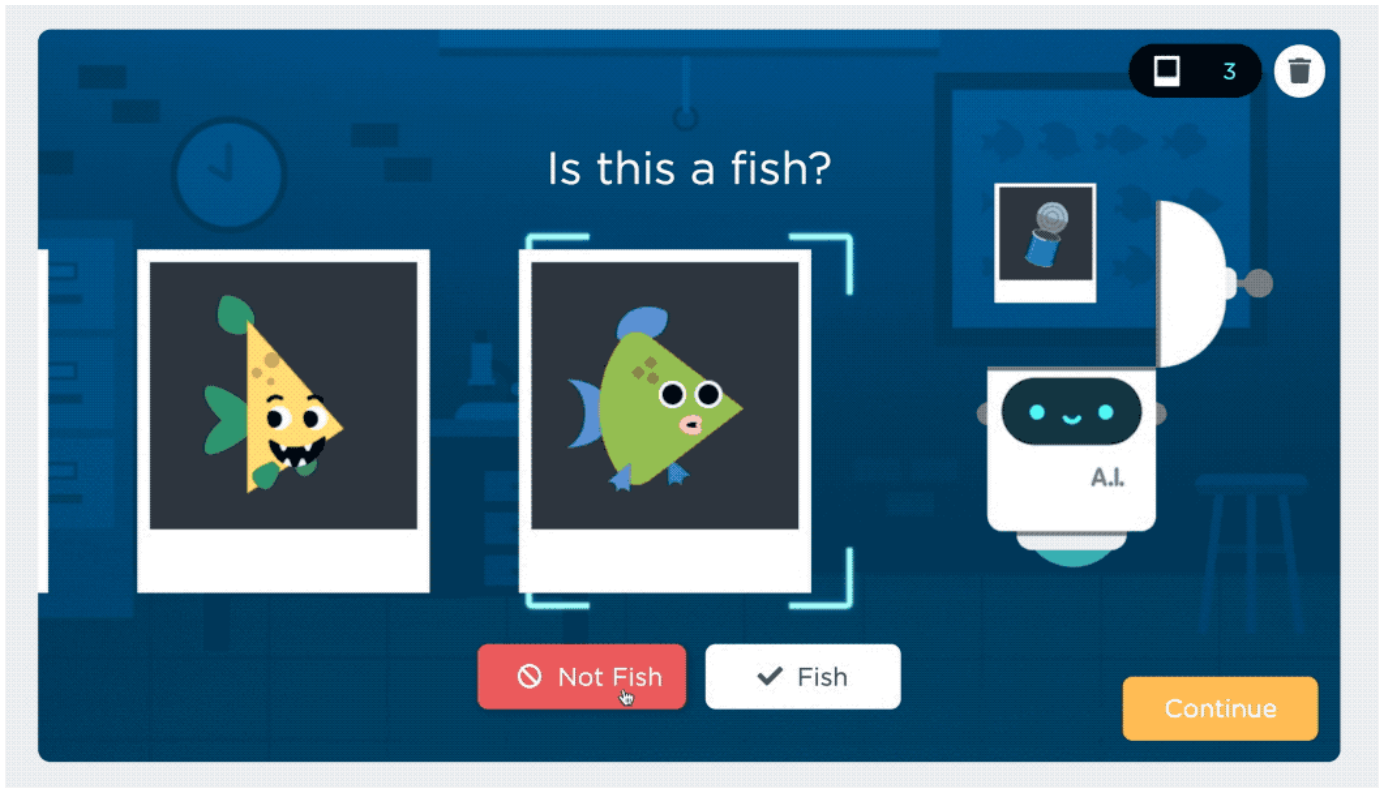}
        \caption{Game: AI For Oceans}
        \label{fig:game}
    \end{subfigure}
    \hfill
    \begin{subfigure}[t]{0.46\linewidth}
        \centering
        \includegraphics[width=\linewidth]{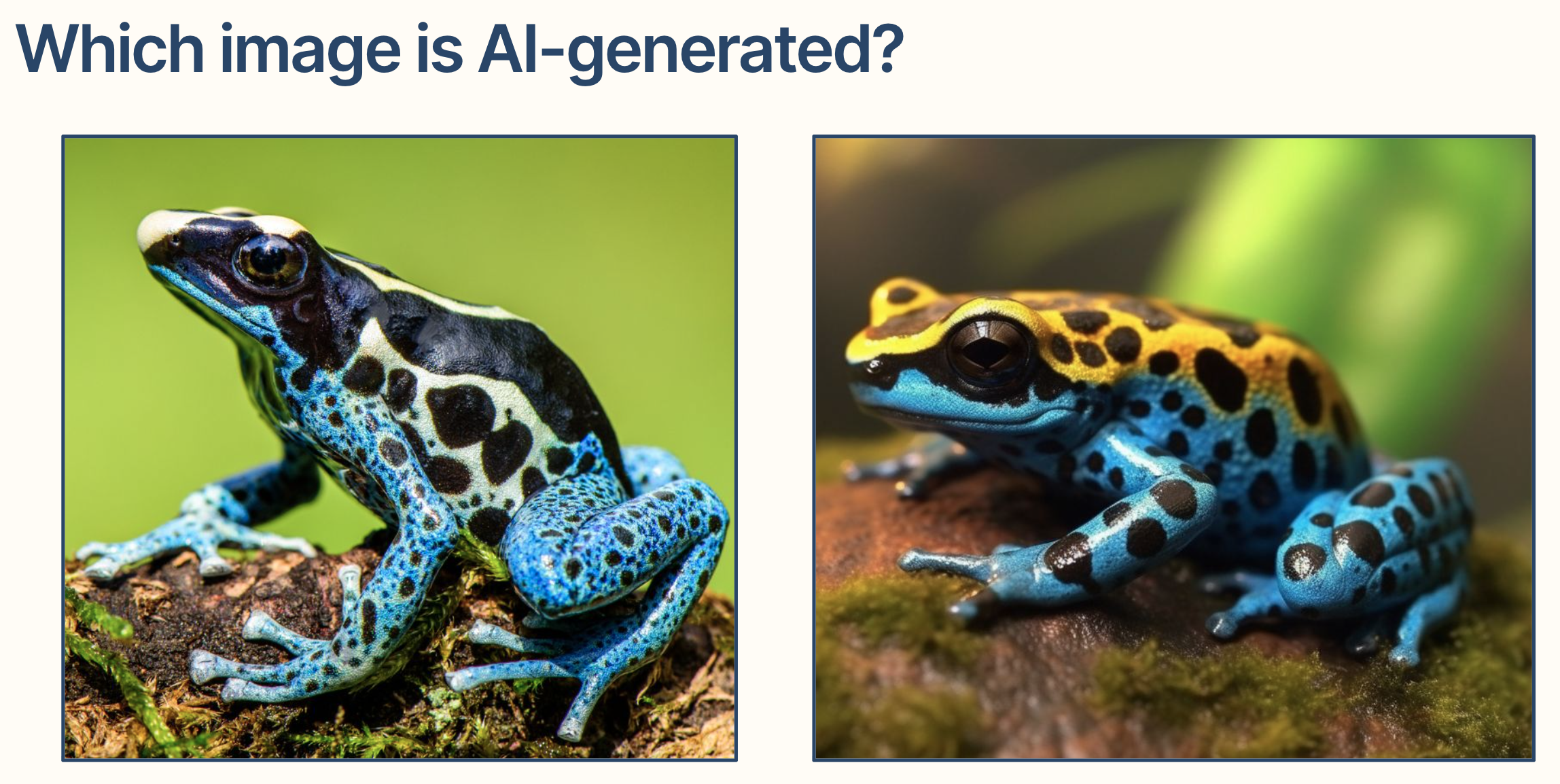}
        \caption{AI Image vs Real Image}
        \label{fig:frog}
    \end{subfigure}
    \caption{Educational Content Design for the Session}
    \Description{Two screenshots used in the AI education session. Left: a screenshot of the ``AI for Oceans'' interactive game from Code.org showing underwater creatures and a machine learning classification interface. Right: two side-by-side frog images, one AI-generated and one real, used in a polling activity where attendees identified which was AI-generated.}
\end{figure}

\textbf{Conceptual Grounding}. The community partners suggested using educational materials suitable for a general audience without prior AI expertise to lower cognitive barriers. The research team assembled a set of materials that consisted of less technical details, such as explanatory videos and interactive games from educational institutions. For example, we adopted "AI for Oceans (See Figure-\ref{fig:game}) \footnote{https://studio.code.org/courses/oceans/units/1/lessons/1/levels/1}" from Code.org, an interactive educational activity for 8-year-old or older students. It uses a game-like scenario to show core concepts like AI, data, training, biases, and the impact of data quality on AI outcomes.

\textbf{Local Relevance}. Local relevance leveraged the community's existing sense of community as a learning resource. Community partners also emphasized making the examples and demonstrations more relevant to attendees' everyday lives and/or family contexts. One of the community partners suggested including deepfake material and discussing how AI can both generate images and replicate voices, given the community partner's knowledge of scams and community members' fears. Therefore, we designed an interactive poll on identifying AI fake images versus real images. In this activity, in addition to animals (See Figure-\ref{fig:frog}), we also used some local landscapes (local stadium) to make the examples more relevant to the audience.

\textbf{Dialogical Learning}. To make sure community members had better learning experiences, we avoided a lecture-style session in which the speaker did most of the talking; instead, we intentionally divided the session into multiple sections that allowed attendees to freely ask questions, share experiences, and discuss their understandings. We treated members as co-producers of AI understanding rather than recipients of expert knowledge.

In addition to the three core decisions, we also ensured: \textbf{Approachable Communication}: making content accessible (at an 8th-grade reading level) and avoiding technical terms; \textbf{Ongoing Engagement}: allowing attendees to provide feedback and follow-up options for future contact; and \textbf{Inclusive Participation}: supporting both in-person and online options, collecting feedback from both paper and digital surveys.

Therefore, this co-created education session ensured its content was community-based, empowering, supporting, and encouraging community members to share experiences, perceptions, and questions \cite{dahl_facilitation_2025}. Both research and community teams approached the session thoughtfully, requiring careful adaptation to a community setting rather than academic classrooms. Our goal was not to focus on technical proficiency but to create a supportive and educational environment that fosters understanding, critical reflection, and agency in continuous learning, because many community members encounter AI primarily through everyday, high-stakes contexts rather than formal technical training.

\titlespacing*{\section}{0pt}{0.2ex}{0.2ex}
\subsection{Co-hosting the Education Session}

The education session was not only the implementation of the education materials but also an organized event that focused on equalizing access to resources for each community member. Event activities were carefully planned and organized to ensure that all attendees were respected, understood, heard, and empowered. Session details were carefully discussed to ensure community members could attend and participate. \textbf{Time}: community partners recommended starting around 5:30 PM to avoid conflicts with the majority of community members' working hours. \textbf{Location}: The session was held at a local community center familiar to community members, rather than at the university, to reduce excessive travel and ensure a safe, familiar space for learning. \textbf{Transportation}: Attendees who needed transportation were also provided with free rides. \textbf{Hybrid Session}: Presentation settings (including both online and offline) ensured that attendees could comfortably and conveniently listen, watch, and discuss. \textbf{Refreshments and Family Support}: Food and drinks were provided, and community members who brought their children were provided with sketchbooks.

All session activities were organized to support reflection, understanding, interaction, and learning. The session began with a sign-in sheet and a pre-session survey to understand attendees' prior experiences with and perceptions of AI. The session was equipped with multiple microphones and a large television screen so everyone could see the slides, and it was recorded. The session concluded with a post-survey focused on changes in understanding of AI. Forty-eight people attended in person, and six people attended virtually via Zoom. The research team and the community partners served as facilitators during the session. A member of the research team presented slides that provided an overview of the work's purpose, the funders, and the aforementioned activities (e.g., videos and interactive activities). Three research assistants helped facilitate the session by observing who had questions, passing around the microphones as needed, assisting with surveys, and providing support to people who needed accessibility assistance. One member of the research team was also online to facilitate online participants. \rev{To ensure inclusive participation, session facilitators actively monitored engagement among attendees throughout the session, encouraging all to speak while also actively identifying and inviting those who had not spoken.}

\section{SESSION FEEDBACK}

\rev{We collected multiple forms of data: surveys at the beginning and end of the session, interactive polling during the session, facilitators' observation notes, and reflections with community partners alongside corresponding notes. We used descriptive analysis for the survey and interactive pooling data and qualitative analysis for the open-ended responses.} 

\textbf{Pre and Post Session Surveys}. \rev{Out of the 54 attendees, 37 completed the pre-survey, and approximately 22–25 completed the post-survey, depending on the question. In the pre-survey, attendees' responses showed broad and generalized concerns about AI, including skepticism about trustworthiness (25 of 35, 71\%), fears of large-scale societal harm (26 of 35, 74\%), distrust of technology companies' ability to self-regulate (26 of 35, 76\%), and worries about privacy (21 of 28, 75\%). In addition, attendees expressed their interest in learning more about AI. In addition to how AI works (28 of 37, 76\%), attendees were interested in how AI impacts various aspects related to their lives, like privacy (24 of 37, 65\%), jobs (23 of 37, 62\%), and everyday life (19 of 37, 51\%). After the session, community members continued to express many of these same concerns. However, we noticed some emotional change between pre- and post-survey results, e.g., curiosity (from 71\%, 24 of 34 in pre-survey, to 50\%, 11 of 22 in post-survey), which likely reflects not a loss of interest but a shift from diffuse curiosity to more concrete and critical engagement. For instance, one attendee said, "It makes me aware of how deceptive AI can be." Post-session survey responses showed greater specificity and grounding in concrete issues discussed during the session. For example, attendees said they learned "having to use AI is inevitable. AI is human-trained," and "How to identify AI images." They also reported increased confidence in their understanding of AI, particularly in distinguishing between types of AI applications and recognizing where human decisions shape AI systems. For instance, the response to ``don't know/not familiar enough to say (about AI)'' dropped from 23\% (8 of 35) in the pre-survey to 0\% (0 of 24) in the post-survey. For example, one participant commented, "The information on different tasks that AI can make easier. The dialogue around the risks lets us know AI is not perfect." These patterns suggested that while the session did not fundamentally alter community members' underlying concerns about AI, it supported a shift from abstract unease to more articulated, specific, and actionable forms of understanding. }

\textbf{Interactive Polling}. The interactive polling began with open-ended questions asking attendees about their first impression of AI. The feedback on the questions surfaced a range of responses, including concerns about job loss (e.g., "Concerned about how AI works, and what qualifications should you have to be able to program it"), surveillance and privacy ("It sounds more like the government gets more control"), and confusion about AI and its application (e.g., "I don't understand what it is yet"). These responses highlighted uneven AI familiarity across attendees and helped facilitators adjust explanations and discussion in real time. Later polling questions introduced concrete examples, including a game-like activity where attendees were asked to determine whether images were AI-generated or not. Attendees correctly identified AI-generated images in 82\% of responses and also explain their rationales. For example, when identifying frogs (See Figure \ref{fig:frog}), one person thought the AI-generated image was real because the background looked like the frog's ``rightful habitat,'' unlike the plain green background in the other image. Another thought the real image was real because of its eyes. Another thought the AI-generated image was AI because it was ``too perfect'', had a ``glow'' to it, and there was something ``off'' with the colors. Also, the observation notes showed that attendees showed strong interest in engaging in such activities because the examples used are based on local items (e.g., local landscape, sports teams). To conclude, the polling activities served as a scaffold for discussion instead of a measurement tool. By prompting attendees to externalize uncertainty and reasoning, polling helped create a shared learning space where participants could build understanding collectively rather than being evaluated on correctness.

\rev{The research team co-designed the educational session, with community partners, and led the analysis of its outcomes. We acknowledge that our interpretations may have been shaped by our own commitments, expertise, and relationships within the project. The senior author, who also served as PI, has maintained decade-long collaborative relationships with both community partners. These relationships informed the co-design process and shaped the trust, assumptions, and shared commitments that structured the work. In addition, the research team's familiarity with AI shaped how session content was framed and interpreted. To support reflexivity and strengthen the credibility of our analysis, we incorporated session findings into our ongoing practice of sharing research progress with community members and partners. Their feedback helped to clarify our assumptions and refine the analysis.}

\section{DISCUSSION}
By presenting a community-informed approach for co-designing an AI education session, we answered "What does AI education for adults outside formal educational systems look like in practice?" and "What does this AI education session reveal about AI literacy at the community level?". Next, we discuss this case with prior work and provide discussions and implications. 

First, this education session suggests the need to reconsider AI literacy as community capacity. Prior work has pointed out that dominant AI literacy frameworks often emphasize technical competencies, individual skill acquisition, or workforce readiness \cite{almatrafi_systematic_2024,ng_AI_2021}, assumptions that may not align with the lived experiences of people outside educational institutions \cite{ko_we_2025}. From our analysis, AI literacy was less about mastering tools or technical mechanisms, but more about developing the capacity to recognize AI in everyday contexts, critically interpret its outputs, articulate concerns and expectations, and make informed decisions. Prior work \cite{chaskin_Building_2001,dillahunt_Development_2025} defines community capacity as the collection and interplay among human, organizational, and social resources that enable a community to recognize and address problems. Our work also aligns with prior CHI and CSCW literature \cite{tang_ai_2025,dahl_facilitation_2025}, calling out the importance of the environment and approach of such sessions, not only the instructional content. We were intentional about creating a supportive, inclusive, and safe space for learning.  We summarize these considerations into two categories: (1) participation-driven environment settings, which cover considerations of time, location, hybrid, transportation, and room arrangement to accommodate community members' everyday life routines, family background, neighborhood condition, financial situation, and accessibility needs; (2) voice-oriented event settings, which include a large variety of activities like surveys, interactive polling, and in-session discussion (also supported by a comprehensive set of technology that could document feedback), that understand, communicate, and empower community members' voices. These two categories build community capacity at the individual access level and the community interpretation level.

Second, moving beyond considering AI literacy as community capacity, we argue that teaching AI in community contexts demands explicit recognition of power dynamics embedded in both technology and education. Prior research has shown that the stance of introductory material on AI affects the audience's long-term judgment of it \cite{reyes-cruz_resisting_2025}. Therefore, the research team also needs a clear position when introducing AI to the community. Rather than framing AI as inherently (or only) beneficial or harmful, the session intentionally surfaced tensions between utility and risk and promise and harm, aligning with CHI and CSCW scholarship, which critiques techno-solutionist narratives and calls for making power, accountability, and harm visible in AI systems \cite{reyes-cruz_resisting_2025}. The education session adopted interactive, context-aware pedagogical strategies inspired by community partners' suggestions, which later proved engaging and effective. Activities like WooClap polls and presentation of AI images of local landmarks sparked rich discussions. Considering that many attendees enjoyed the informal discussion during the session, our interactive approach also supports prior work arguing for valuing conversational, embodied, and collective forms of engagement alongside quantitative interaction metrics (such as the number of survey responses) \cite{gautam_Dynamic_2024}. 

Third, these insights suggest several implications for future research and practice in community-centered AI education. (1) Future work should move beyond the technical skill-centric implications of AI literacy and develop frameworks, measures, and interventions that are explicitly grounded in specific populations, especially those who have not been the primary focus in prior work. This should include examining AI literacy as a collective and community-level capacity (rather than solely an individual competency) \cite{dillahunt_Development_2025, dillahunt_Measure_2024,hui_Was_2024}, and designing evaluation methods that capture critical awareness, sense-making, and decision-making (rather than technical mastery alone). (2) Researchers and practitioners should treat the creation of supportive, inclusive, and safe learning environments \cite{dahl_facilitation_2025} as a core design problem rather than a logistical afterthought, systematically studying how participation-driven and voice-oriented settings shape engagement, trust, and learning outcomes over time. (3) Future AI education initiatives should continue to experiment with pedagogical approaches that foreground power, accountability, and local relevance, using interactive and situated strategies to surface tensions and foster communication (rather than promote simplified narratives of AI).

\begin{acks}
\rev{We thank our research assistants, Feier Sophie Su, Tianyu Hu, Pari Dar, and Jay Davis, and all community participants. This work was supported by the National Science Foundation (No. 2427332). We used LLM-based tools for grammar editing only; all conceptual framing, analyses, interpretations, findings, and references were developed and verified by the authors.}
\end{acks}

\bibliographystyle{ACM-Reference-Format}
\bibliography{Reference/sophie,Reference/ReDDDot,Reference/timebank,Reference/SophieZotero,Reference/references}

\newpage
\clearpage
\onecolumn

\appendix
\section{APPENDIX: Figures}

\section{APPENDIX: Survey Question Examples}
\begin{table}[H]
\centering
\textbf{Q1.} Please state your level of agreement with the following statements about the future of AI and its impact on society.

\smallskip
\noindent{\footnotesize\textit{Scale: SA = Strongly Agree, A = Agree, SWA = Somewhat Agree, SWD = Somewhat Disagree, D = Disagree, SD = Strongly Disagree, DK = Don't Know / Not Familiar Enough to Say}}
\smallskip

\begin{tabular}{@{}p{10cm}ccccccc@{}}
\toprule
 & \footnotesize SA & \footnotesize A & \footnotesize SWA & \footnotesize SWD & \footnotesize D & \footnotesize SD & \footnotesize DK \\
\midrule
AI can increase efficiency and accuracy in tasks     & & & & & & & \\
AI can offer convenience and save time               & & & & & & & \\
AI may violate privacy concerns                      & & & & & & & \\
I trust AI to make reliable decisions                & & & & & & & \\
I trust AI to be used ethically                      & & & & & & & \\
AI may perpetuate bias and discrimination            & & & & & & & \\
\bottomrule
\end{tabular}
\end{table}

\begin{table}[H]
\centering
\textbf{Q2.} Please state your level of agreement with the following statements.

\smallskip
\noindent{\footnotesize\textit{Scale: SA = Strongly Agree, A = Agree, SWA = Somewhat Agree, SWD = Somewhat Disagree, D = Disagree, SD = Strongly Disagree, DK = Don't Know / Not Familiar Enough to Say}}
\smallskip

\begin{tabular}{@{}p{10cm}ccccccc@{}}
\toprule
 & \footnotesize SA & \footnotesize A & \footnotesize SWA & \footnotesize SWD & \footnotesize D & \footnotesize SD & \footnotesize DK \\
\midrule
 I am concerned about the reliability of the information provided by AI & & & & & & & \\
 I am worried that machines with AI could eventually pose a threat to the existence of the human race & & & & & & & \\
 Technology company executives can't be trusted to self-regulate the AI industry & & & & & & & \\
\bottomrule
\end{tabular}
\end{table}
\end{document}